\documentclass[prb,preprint]{revtex4-1} 
\usepackage{amsmath}  
\usepackage{amsfonts} 
\usepackage{graphicx} 
\begin{document}
\title{SEARCHING FOR A HIGHLY UNLIKELY FRAME DEPENDENT SPEED OF LIGHT USING A ONE-WAY TEST}
\author{Md. Farid Ahmed$^{1}$}
\email{mdfarid@yorku.ca; mdfaridyorku@gmail.com} 
\author{Brendan M. Quine$^{1, 2}$}
\author{Spiros Pagiatakis$^{1, 2}$}
\author{A. D. Stauffer$^{2}$}
\affiliation{$^{1}$Department of Earth and Space Science, York University, 4700 Keele Street, Toronto, Ontario, Canada-M3J 1P3.}
\affiliation{$^{2}$Department of Physics and Astronomy, York University, 4700 Keele Street, Toronto, Ontario, Canada-M3J 1P3.}
\date{\today}
\begin{abstract}
At first blush, what appears to be a purely physical question to measure any velocity: how to measure the velocity on a one-way trip? However, due to the debates of the clock-synchronisation and the successes of Special Relativity (SR), searching of the possibility of one-way speed of light measurement did not receive wider attention since the declaration of the constancy of the speed of light in vacuum by Maxwell's Electrodynamics in 1864. However, our analysis suggests that the debates of the clock synchronization are insignificant when one uses a one-way experiment to test the isotropy of the speed of light - the fundamental postulate of SR. Searching of the possibility of one-way speed of light to test SR is introduced by reviewing recent one-way tests.  
\end{abstract}
\maketitle 
\section{Introduction} In 1864, J. C. Maxwell had set forth the problem of ether drift by the declaration of electrodynamics and despaired of a solution by experiment \cite{Sw1}. Over the past 150 years, a significant number of experimental investigations have been performed in different ways on the earth and on the space stations \cite{AB2}. These widely used experiments can be divided into two types based on the Robertson-Mansoury-Sexl (RMS) \cite{RB3, MS4, MS5, MS6} test theory: (a) one-way [or first order $(\frac{v}{c})$] test; (b) two-way [or second order $ (\frac{v^{2}}{c^{2}}) $] test, where $ v $ is the speed of the laboratory and $ c $ is the speed of light in vacuum. These tests are mostly based on a traditional approach using two-way experiments \cite{AB2, KM7} where one tries to find an upper limit of the second order variation which is very small compared with the speed of light $ c $. In principle, a first order effect is $ 10^{4} $ times higher than a second order effect [assuming $ v=30 $km/s, the orbital speed of the earth, and $c=3\times 10^{5} $ km/s, the speed of light in vacuum]. However, one-way speed of light measurement did not receive wider attention because of the debates of the clock synchronization and successes of the SR. \\ \\
Due to the present challenge in the formulation of a quantum theory of gravity \cite{KS8, AME9, GP10}, the fundamental postulate of SR - the constancy of the speed of light - is now under further scrutiny in many different ways \cite{CH11}. The methods and techniques of measurement have been extended and improved with the improvements of technology with time and our dependence on their reliability has increased. Therefore, researchers repeat tests of the fundamental postulate of any established theory over time to get more reliable result. Here, we attempt to indicate a direction for further investigation of the possibility of the one-way speed of light measurement to test the isotropy of the speed of light using the experiments that directly used the one-way speed of light in the recent years. A treatment of the motion of the earth with respect to a preferred frame is considered for present analysis.
\section{Motion of the earth and isotropy test}
\begin{figure}[h!]
\centering
\includegraphics[width=5in]{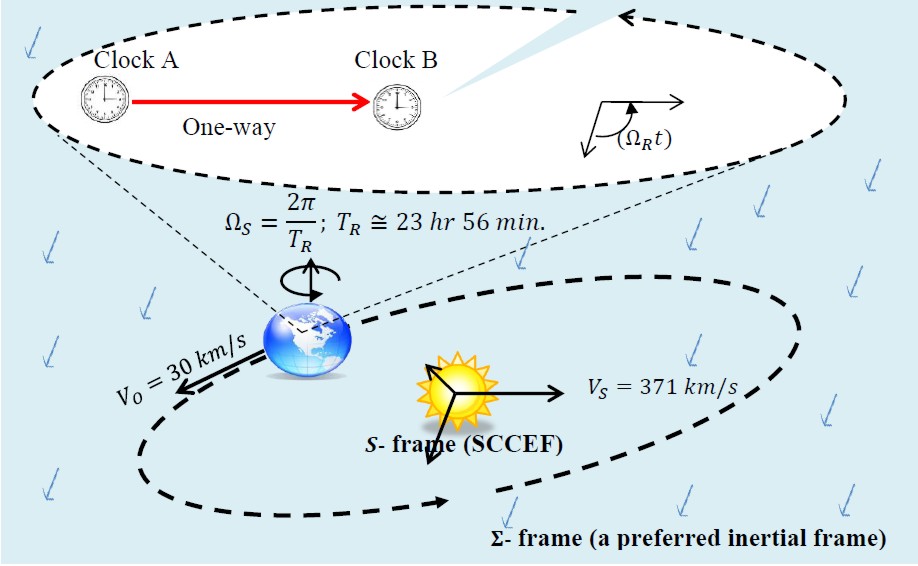}
\caption{Schematic of the SCCEF ($S$-frame) with speed $ \geq 371 $ km/s relative to the CMB ($ \Sigma $-frame) with an idea of a conventional one-way speed of light test.}
\label{fig:shematic1}
\end{figure}

\begin{figure}[h!]
\centering
\includegraphics[width=5in]{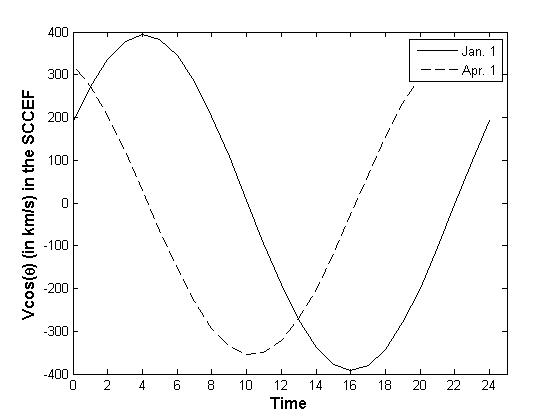}
\caption{An illustration of the diurnal variation and phase change of velocity of the laboratory relative to a rest frame (SCCEF)}
\label{figure:shapechange2}
\end{figure}

\begin{figure}[h!]
\centering
\includegraphics[width=5in]{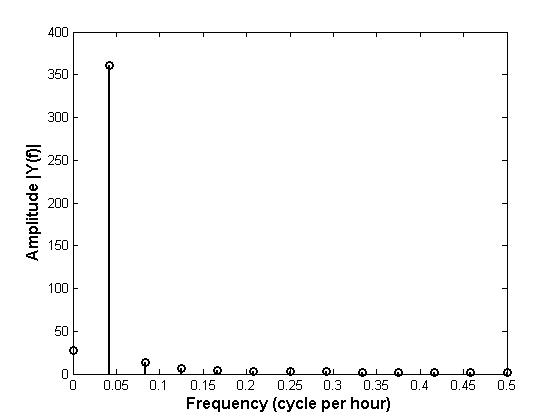}
\caption{An illustration of the Fourier analysis of Fig. \ref{figure:shapechange2} following \cite{JE34}.}
\label{figure:fourier3}
\end{figure}

We admit that motion of the earth does occur. Then we immediately recognize that the concept of velocity of the earth plays a key role in our description and understanding of the isotropy of speed of light tests in a terrestrial experiment \cite{SH36}. \\ \\
Different co-ordinate transformations are proposed to describe space-time \cite{SP12}. Among them, the well-known Galilean transformations based on a universal time which leads to the anisotropy of the speed of light while Lorentz transformation is based on the constancy of the speed of light. RMS transformations (or test theory) assume that the speed of light is isotropic in the rest frame but needs to be tested in the moving frame. In order to give an idea of RMS test theories compared with Galilean (Newtonian) and Lorentz transformations, we introduce Fig. \ref{fig:shematic1}. Let us consider two inertial reference frames where $ \Sigma $ is the hypothetical rest frame [the Cosmic Microwave Background (CMB) is the best candidate with experimental proof] and $ S $ [the Sun-Centered Celestial Equatorial Frame (SCCEF)] is moving at a uniform velocity $  v$ along the hypothetical rest frame. The CMB and the SCCEF have been elaborated in detail in \cite{AB2, HE13,KO14, MU15,SA16, TO17, WO18} to explain different tests of the Lorentz invariance. The SCCEF is the frame in which the Sun is at the centre, and is inertial relative to the CMB frame to first order. In addition, we introduce a brief idea of a conventional one-way test where two clocks are required (Clock A and clock B) as shown in Fig. \ref{fig:shematic1}. 
According to RMS, the speed of light is isotropic in this rest frame $ \Sigma $. Following the reports \cite{MS4, WI19}, we can parameterize the orientation dependence of the one-way velocity of light as measured in the inertial $S$-frame in which the laboratory is at rest as follows: 
\begin{equation}
\label{eq:rms1}
c(t)\cong c \left[ 1+ A \frac{v}{c} \cos \left( \Omega_{S}t \right) \right]; ~~~~~~~~~~~~~ 0\leq t \leq 23~hr~56~min 
\end{equation}
where $c$ is the constant speed of light in the $ \Sigma $-frame, $\left(  \Omega_{S}t \right)  $ is the angle between the direction of light propagation and the velocity vector of the $S$-frame relative to $ \Sigma $-frame and $\Omega_{S}=\frac{2\pi}{T_{S}}  $; $ T_{S}=23~hr~56~min $. If SR is valid then, $A=0$. A one-way experiment can set upper limits on $A$. \\ \\
We derived the time dependent components of the velocity of the laboratory along the direction of the light propagation in the paper \cite{AB2} assuming the Cosmic Microwave Background (CMB) is the rest frame of the universe. This derivation can help us to understand the shape of the change of velocity of the laboratory relative to the rest frame. For example, following the propagation direction of light in our laboratory in the East-West direction, we derive the time dependent component of the velocity of the laboratory relative to the rest frame along the direction of light propagation as follows \cite{AB2}:
\begin{equation}
\label{eq:eastwest2}
\begin{split}
v(t)=[\sin( \Omega_{S}t)][V_{O}\sin (\Omega_{O}t)+V_{R}\sin (\Omega_{S}t)-V_{S}\cos (\alpha) \cos (\delta)] \\ +[\cos (\Omega_{S}t)][V_{O}\cos (\varepsilon) \cos (\Omega_{O}t)+V_{R}\cos (\Omega_{S}t)+V_{S}\sin (\alpha) \cos (\delta)] 
\end{split}
\end{equation}
where $ V_{S}=371 $ km/s is the velocity of the solar system towards $ [(\alpha,\delta)=(168^{\circ}),-7.22^{\circ}] $ relative to the CMB, where $ \alpha =$ right ascension and $ \delta= $ declination; the angle, $ \varepsilon =23.4^{\circ} $ \cite{SM20}. $ V_{O} $ and $ V_{R} $ are the orbital and rotational velocities of the earth respectively. $ \Omega_{O} $ is the orbital frequency with respect to a fixed star. \\ \\
Figs. \ref{figure:shapechange2}  and \ref{figure:fourier3} illustrate the shape changes of velocity of the laboratory relative to the rest frame that can help us to understand the one-way isotropy test according to a Galilean transformation.
\section{Recent one-way tests}
\begin{figure}[h!]
\centering
\includegraphics[width=5in]{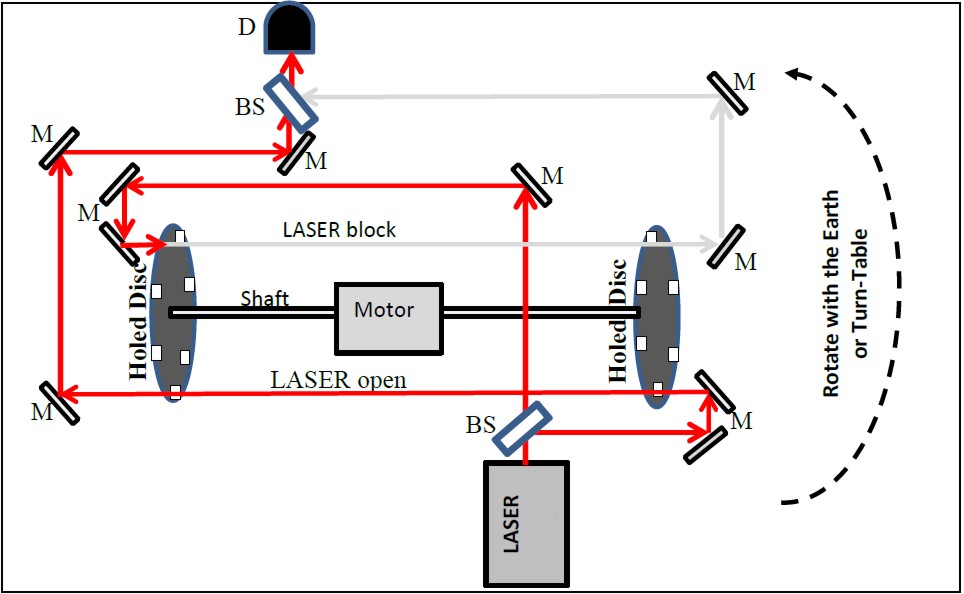}
\caption{Block diagram of Fizeau-type-coupled-slotted-discs with possible future improvements using only one detector, to test one-way isotropy of light. [According to this improvement, the detector is able to receive only one direction light at a time. Here, D is Photodiode; M is Mirror; BS is Beam-Splitters].}
\label{figure:block3}
\end{figure}

\begin{figure}[h!]
\centering
\includegraphics[width=5in]{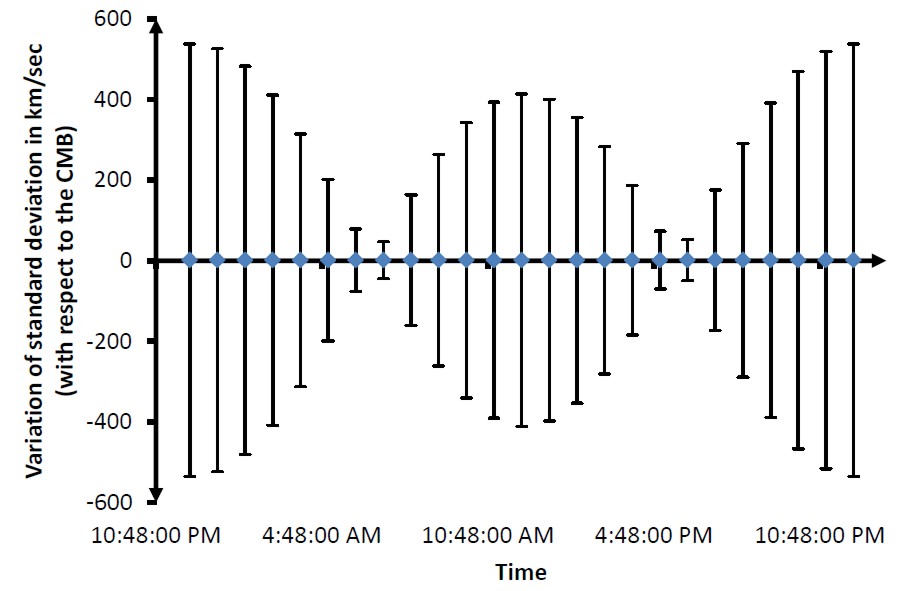}
\caption{Expected variation in the standard deviation of results according Fig. 2 if the speed of light follows a Galilean transformation.}
\label{figure:diur4}
\end{figure}

By examining recent experiments that directly use one-way light, we continue our discussion of the possibility of one-way speed of light to test the isotropy of SR. \\ \\
In 2009, the paper \cite{GE21} reported a one-way experiment using a time of flight technique. This experiment used a light beam from a He-Ne laser. The beam was amplitude modulated and was directed onto a detector. The response was converted to an electrical signal. The phase-difference between the transmitted and the received signals was measured. The paper \cite{GE21} presents the average value of the one-way speed of light of two measurements which is $ 3.009\times 10^{8} $ m/s. \\ \\
In 2008, another report \cite{CA21} indicated similar type of a one-way speed of light measurement of the report \cite{GE21} using two arrangements with identical equipment of stable frequency sources and pulse generators that make simultaneous one-way speed of light measurements from clock A to clock B and in the opposite direction. The comparison of the two opposite responses was proposed. Also the report \cite{CA21} indicates that the one way light speed to an accuracy around 1 part in $ 10^{6} $. \\ \\
In order to test one-way isotropy of the speed of light, we would like to indicate that this experiment should be performed for a period of 24-hours when the earth makes a complete rotation. Then, searching of the possibility of the consequences of the speed of earth in terms of one-way speed of light measurement using this experiment can be performed assuming a Galilean transformation as follows:
\begin{enumerate}
  \item One can plot the responses with time of the day of measurement and look for any diurnal variation as illustrated in Fig. \ref{figure:shapechange2}. In addition, Fourier transformation of the data can give the idea about any meaningful frequency that is consistent with the earth's rotation as simulated in Fig. \ref{figure:fourier3}.   
  \item With the existence of any meaningful diurnal variation, one can repeat the measurement in different seasons of the year and look for any meaningful phase change due to the orbital motion of the earth as illustrated in Fig. \ref{figure:shapechange2} using simulation in different seasons of the year.
  \item One can do this measurement simultaneously with the light propagations in opposite directions as indicated in the report \cite{CA21}. The plot of the outcomes following numbering 1 would be in opposite phase if the speed of the earth can be measured in terms of the speed of light. A normalization approach is derived and described in the report \cite{AB27} and used in the report \cite{AB28}. This approach can make the two opposite responses identical for precise comparison to look for any opposite phase in the responses. 
\end{enumerate}
Fizeau's gear-wheel method for measuring the speed of light \cite{DI22} was adopted and improved in the reports \cite{MA23, MA24, MA25} to measure the hypothetical variation of the one-way speed of light and reported a controversial result \cite{MA26}. This experiment was proposed to re-consider in the reports \cite{AB2, AB27} in response to the reports \cite{MT33, RA34}. Detailed descriptions of the experimental apparatus, different challenges and an interpretation of the experiment have been presented elsewhere \cite{AB27, AB28}. Recently in 2013, report \cite{AB28} presents the outcome of Fizeau's gear-wheel method for measuring the one-way speed of light that is consistent with SR. \\ \\
The interpretation derived in \cite{AB27} suggests that the test using two detectors is insensitive to measure any velocity of the Earth toward the direction parallel to the spin-axis of the Earth. Taking into account this limitation as well as other challenges to use two-detectors, we propose a new approach of the test of the one-way isotropy of the speed of light using one detector in a Fizeau-type-coupled-slotted-discs experiment in this section. This will overcome some of the limitations of our present experiment \cite{AB27, AB28}. \\ \\
A block diagram of our proposed future improvement to test the isotropy of the one-way speed of light using a Fizeau-type-coupled-slotted-discs is shown in Fig. \ref{figure:block3}. In this improvement, we consider the response of the detector to only one direction of the light at a time. When the light travelling in a direction is received by the detector, the light in the opposite direction is blocked by the disc and vice versa. 
The data collected in the course of a 24-hour period will not present any significant change if the speed of light follows SR. However, if SR is not valid then we will observe a significant regularity of variation in the collected data as simulated in Fig. \ref{figure:diur4} following in Fig. \ref{figure:shapechange2}. If special relativity is valid then there will not be any diurnal variation. 
The experiment on the turn-table would be the best setup as one can perform this test within a short period of time. 
\section{Discussions}
The one-way velocity of light depends on the synchronization parameter \cite{MS4, AB29}. Report \cite{WI19} indicates that experiments which test the isotropy in one-way or two-way have observables that depend on test functions but not on the synchronization procedure. Therefore, the consideration of clock synchronization is irrelevant for a one-way speed of light test if one expresses the results in terms of physically measurable quantities \cite{WI19}. It is possible to measure a one-way speed of light \cite{SP12, SP30}. \\ \\
The authors of the reports \cite{SH31, ED32} emphasized that the "one-way speed of light has so far never been examined in experiments".  Contrary to this quotation, however, it is already noted in the papers \cite{AB2, WI19} that there are experimental investigations of the one-way speed of light. The controversy about the results on the limits of the isotropy of the one-way speed of light from NASA-experiments \cite{AB2, WI19} and the regularity in the variations of the reported results of the isotropy of the one-way speed of light in different time periods of the GRAAL facility of the European Synchrotron Radiation Facility (ESRF) in Grenoble \cite{AB2, ZH35, GA36,GA37, GA38} remain unclear and needs further investigations by one-way experiments. \\ \\
The previous one-way experiments noted above are not repeatable in simple forms in different laboratories as one can do with Michelson-Morley type two-way tests. In order to look for a one-way experiment to test the isotropy of the speed of light which is repeatable in a simple form in any laboratory and also, in order to ensure the results are clearly valid, we are proposing an improvement to our previous approach in \cite{AB27, AB28} using one detector. In addition we indicate the possibility of a one-way speed of light isotropy test in the work reported in the papers \cite{GE21, CA21}.   
\section{Conclusion}
Based on above discussions, we would like to emphasize at this point that a general point of view of the fundamental postulate of SR, i.e., the "one-way" constancy of the speed of light is not accepted unambiguously and is still controversial \cite{AB2, SH31, ED32, MT33, RA34, ZH35, GA36, GA37, GA38, BO39}, and needs further investigations from one-way experiments. \\ \\
We would like to emphasize that one-way tests should not be dismissed due to the perceived problem of clock synchronization \cite{SP12, SJ40, PO41}. It has been shown how the consequences of a one-way speed of light can be used to perform isotropy test assuming the light follows a Galilean transformation. A one-way speed of light measurement in terms of physically measurable quantities for a period of 24-hours with graphical representation can indicate the following phenomena if SR is invalid:
\begin{enumerate}
  \item A diurnal variation   
  \item Spike at the frequency of $(1/24)$ cycle /hour in Fourier spectral analysis.  
  \item Phase shifts for the measurement in different seasons.
  \item Opposite phase in the responses of the two simultaneous measurements. 
\end{enumerate}
A magnitude of the one-way velocity of light can be challenging due to the debates of the clock-synchronisation. However, we conclude that as long as one performs a one-way speed of light measurement for a period of 24-hours and present the outcomes in terms of physically measurable quantities, one-way isotropy test of SR can be performed.      
\section*{Acknowledgements}
This work is supported in part by the National Science and Engineering Research Council, Thoth Technology Inc. and York University, Canada.      


\begin{thebibliography}{99}
\bibitem{Sw1}
Swenson L. S., Jr. \textit{The ethereal aether: a history of the Michelson-Morley-Miller aether-drift experiments}, (1880 - 1930, University of Texas Press, USA, p. 234) (1972).
\bibitem{AB2}     
M. F. Ahmed, B. M. Quine, S. Sargoytchev, and A. D. Stauffer, Indian J. Phys. \textbf{86}, 835 (2012 [arXiv:1011.1318]. 
\bibitem{RB3}
H.P. Robertson, Rev. Mod. Phys. \textbf{21}, 378 (1949).
\bibitem{MS4}   
R. Mansouri and R.U. Sexl, Gen. Rel. Grav. \textbf{8}, 497 (1977).
\bibitem{MS5}
R. Mansouri and R.U. Sexl, Gen. Rel. Grav. \textbf{8}, 515 (1977).
\bibitem{MS6}
R. Mansouri and R.U. Sexl, Gen. Rel. Grav. \textbf{8}, 809 (1977).
\bibitem{KM7}
T. P. Krisher, L. Maleki, G. F. Lutes, L. E. Primas, R. T. Logan, and J. D. Anderson, C. M. \indent ~Will, Phys. Rev. D, \textbf{42}, 731(1990).
\bibitem{KS8}
V.A. Kosteleck$\acute{y}$  and S. Samuel, Phys. Rev. D \textbf{39}, 683 (1989).
\bibitem{AME9}
G. Amelino-Camelia, J. Ellis, N.E. Mavromatos, D.V. Nanopoulos and S. Sarkar, Nature \textbf{393}, \indent ~763 (1998). 
\bibitem{GP10}
R. Gambini and J. Pullin, Phys. Rev. D \textbf{59}, 124021 (1999).
\bibitem{CH11}
A. Cho, Science, \textbf{307}, 866 (2005). 
\bibitem{SH36} 
R. S. Shankland, S.W. McCuskey, F. C. Leone and G. Kuerti, Rev. Mod. Phys., \textbf{27}, 167 (1955).
\bibitem{SP12}
G. Spavieri, Phys. Rev. A. \textbf{34}, 1708 (1986).
\bibitem{HE13}
S. Herrmann, A. Senger, E. Kovalchuk, H. M$ \ddot{u} $ller and A. Peters, Lect. Notes 
Phys. \textbf{702}, 385 (Springer-Verlag Berlin Heidelberg)(2006). 
\bibitem{KO14}      
V.A. Kosteleck$\acute{y}$ and M. Mewes, Phys. Rev. D \textbf{66}, 056005 (2002).
\bibitem{MU15}
H. M$ \ddot{u} $ller, S. Herrmann, C. Braxmaier, S. Schiller and A. Peters, App. Phys. B: Lasers and Optics \textbf{77}, 719 (2003).
\bibitem{SA16}
G. Saathoff, S. Karpuk, U. Eisenbarth, G. Huber, S. Krohn,  Horta R. Munoz, S. Reinhardt, D. Schwalm, A. Wolf and G. Gwinner, Phys. Rev. Lett. \textbf{91}, 190403 (2003).
\bibitem{TO17}
M. E. Tobar, P. L. Stanwix, M. Susli,  P. Wolf, C.R. Locke and E. N. Ivano, Lect. Notes Phys. \textbf{702}, 416 (Springer-Verlag Berlin Heidelberg)(2006).
\bibitem{WO18}
P. Wolf, S. Bize, M.E. Tobar,  F. Chapelet, A. Clairon, A.N. Luiten and G. Santarelli, Lect. Notes Phys. \textbf{702}, 451 (Springer-Verlag Berlin Heidelberg)(2006).
\bibitem{WI19} 
C. M. Will, Phys. Rev. D \textbf{45}, 403 (1992).
\bibitem{SM20}
G. F. Smoot,  M. V. Gorenstein and R. A. Muller, Phys. Rev. Lett. \textbf{39}, 898 (1977). 
\bibitem{GE21}
E. D. Greaves, An Michel Rodríguez and J. Ruiz-Camacho, Am. J. Phys. \textbf{77}, 10 (2009).
\bibitem{CA21}
J. E. Carroll, Measuring a one way light speed, Cambridge University, Submission for PIRT (2008).
\bibitem{DI22}
R. W. \textit{Ditchburn Light}, (Blackie and Son Limited, London, p. 411) (1963).
\bibitem{MA23}
S. Marinov, Gen. Rel. Grav. \textbf{12}, 57 (1980).  
\bibitem{MA24}  
S. Marinov, Found. of Phys. \textbf{8}, 137 (1978). 
\bibitem{MA25}
S. Marinov, Progress in Physics, \textbf{1}, 31 (2007).
\bibitem{MA26} 
J. Maddox, Nature \textbf{346}, 103 (1990); A.J. Bunting, Nature \textbf{346}, 694 (1990); J. Maddox, Nature \textbf{311}, 399 (1984); H. Aspden, Nature \textbf{318}, 317 (1985); J. Maddox, Nature \textbf{316}, 209 (1985); Nature \textbf{300}, 566 (1982); J. Timno, Nature \textbf{317}, 772 (1985).
\bibitem{AB27}  
M. F. Ahmed, B. M. Quine, S. Sargoytchev, and A. D. Stauffer, [arXiv:1103.6086v3] (2011). 
\bibitem{AB28} 
M. F. Ahmed, B. M. Quine, S. Pagiatakis, A. D. Stauffer, [arXiv:1310.1171] (2013).
\bibitem{AB29} 
C. L$ \ddot{a} $mmerzahl, Lect. Notes Phys. \textbf{702}, 349 (Springer-Verlag Berlin Heidelberg)(2006). 
\bibitem{SP30}
G. Spavieri, Eur. Phys. J. D. \textbf{66}, 1 (2012).
\bibitem{SH31}  
J. Q. Shen, Int. J. Theor. Phys. \textbf{47}, 751 (2008).
\bibitem{ED32}    
W. F. Edwards, Am. J. Phys. \textbf{31}, 482(1963).
\bibitem{MT33}
A. K. A. Maciel and J. Tiomno, Phys. Rev. Lett. \textbf{55}, 143  (1985).
\bibitem{RA34}  
V. V. Ragulsky, Phys. Lett. A \textbf{235}, 125 (1997).
\bibitem{ZH35}  
L. L. Zhou and B-Q Ma, Mod. Phys. Lett. A, \textbf{25}, 2489  (2010) [arXiv:1009.1675]. 
\bibitem{GA36}
V. G. Gurzadyan \textit{et al.} [19 additional authors], Mod. Phys. Lett. A, \textbf{20}, 813 (2005). 
\bibitem{GA37}
V. G. Gurzadyan \textit{et al.} [33 additional authors], II Nuovo Cimento B, \textbf{122}, 515 (2007).
\bibitem{GA38}
V. G. Gurzadyan \textit{et al.} [34 additional authors], (Proc. 12th M.Grossmann Meeting on General Relativity, v.B, p.1495, World Sci.), (2012)[arXiv:1004.2867].
\bibitem{BO39}
J. -P.  Bocquet et al. [37 additional authors], Phys. Rev. Lett., \textbf{104}, 241601 (2010).
\bibitem{SJ40}
T. S. Sj$ \ddot{o} $din and M. F. Podlaha, Lett. Nuovo Cimento, 31, 433 (1982);
\bibitem{PO41}  
M. F. Podlaha, Lett. Nuovo Cimento 28, 216 (1980) 
\bibitem{JE34}
G. M. Jenkins and D. G. Watts, \textit{Spectral Analysis and its applications}, (Holden-Day, Inc., USA)(1968).
\end{thebibliography}
\end{document}